\documentstyle[12pt]{article}

0

\begin{document}

\hspace{12cm} hep-th/9706131

\begin{center}

{}~\vfill

{\large \bf On the geometry of soft breaking terms and $N=1$ 
superpotentials }

\end{center}

\vspace{10 mm}

\begin{center}

{\bf C\'{e}sar G\'{o}mez} 
  
\vspace{7 mm}

{\em Instituto de Matem\'{a}ticas y F\'{\i}sica Fundamental,
CSIC, \protect \\ Serrano 123, 28006 Madrid, Spain}  

\vspace{5 mm}
  
{\em and}
  
\vspace{5 mm}

{\em Instituto de F\'{\i}sica Te\'{o}rica (UAM-CSIC), 
C-XVI, Universidad Aut\'{o}noma de Madrid, Cantoblanco 28049,
Madrid, Spain}

\vspace{5mm}

\end{center}      

\vspace{15mm}
   
\begin{abstract}
We describe, in the context of $M$ theory on elliptically 
fibered Calabi Yau fourfolds, the change of variables that 
allow us to pass from the $U(1)$ invariant elliptic fibration 
to the one decribing the uncompactified four dimensional limit, 
where the $U(1)$ symmetry is broken. These changes of variables 
are the analog to the ones used to derive from the Atiyah-Hitchin 
space the complex structure of 
the Seiberg-Witten solution for $N=2$ pure supersymmetric 
Yang Mills. The connection between these changes of variables 
and the recently introduced rotation of branes is discussed.
\end{abstract}

\pagebreak

\section{Introductory remarks}

One of the most interesting aspects of the celebrated 
Seiberg-Witten solution of $N=2$ SSYM \cite{SW,SW2} 
theories is the way this solution makes operative `t Hooft's 
confinement program based on the so called "abelian proyection" 
gauge \cite{ap}. The beauty of the SW solution for the 
supersymmetric case is that it allow us to pass from the simple 
"parametrization of ignorance", summarized in the soft breaking 
term $\epsilon \hbox{tr} \phi^2$, to the confinement dynamics of $N=1$ SSYM theory as 
described in terms of monopole condensation. In more concrete 
terms the solution of $N=2$ SSYM constitutes an 
effective procedure to pass from the a priori "blindness" - to 
concrete, $N=1$ dynamics- soft breaking term, to a well 
defined $N=1$ superpotential with the set of minima 
predicted by direct computation of $Tr(-1)^F$ \cite{Wind}, 
and gaugino condensates \cite{NSVZ,Shif,A,KS,CG}. How this 
marvelous thing is happening?. The answer is of course 
enclosed in the rich geometry of the $N=2$ solution, 
more specifically in the types of singularities of the 
elliptic fibration defining the solution. The essential 
point consists in giving physical meaning to the SW 
curve itself. This can be done, in principle, in three 
different ways. One by considering field theory limits 
of string compactifications \cite{KKLMV,KLMVW,GHL}, 
another by working 
out brane configurations \cite{HW}- \cite{Wlast} and, finally, by considering the 
compactified theory in three dimensions \cite{SW3d}.

 The key for this last approach is of course based on the 
main result of reference \cite{SW3d}, where it is proved that 
in terms of complex structure the moduli of the $N=4$ three 
dimensional supersymmetric theory is the same of that of the 
$N=2$ in four dimensions, the difference being determined by 
the geometry, namely the volume of the elliptic fiber becomes 
zero in the four dimensional limit (a typical F-theory limit). 
Most of the comments in this note will refer to issues that 
can be beautifully understood using brane configuration 
technology. However, we feel that a different point of view, 
based in the decompactification picture, can be of some value.

\section{Kodaira singularities, soft breaking terms and 
hyperk\"ahler structure}
Let us consider a generic elliptic fibration \cite{Kodaira}:
\begin{equation}
\phi: V \rightarrow \Delta,
\label{1}
\end{equation}
with a set of points $a_i$ where the fiber becomes singular. 
Let assume $\Delta$ compact and of genus zero. The different 
singularities in $\Delta$ are characterized in terms of Dynkin 
diagrams and related monodromy matrices for the elliptic 
modulus. For the $A$ and $D$ cases we have
\begin{eqnarray}
A_{n-1} & = & \left( \begin{array}{cc} 1 & n  \\
			 0 & 1 
	\end{array} \right), \nonumber \\
D_n     & = & \left( \begin{array}{cc} -1 & 4-n  \\
			 0 & -1 
	\end{array} \right).     
\label{2}
\end{eqnarray}
The singular fiber ${\cal C}_i$ at the point $a_i$ is characterized 
by a set of irreducible components satisfying $\Theta_{ij}^2=-2$,
 and 
with the intersection matrix the corresponding affine 
Dynkin diagram:
\begin{equation}
{\cal C}_i = \sum_{j} \Theta_{ij}.
\label{3}
\end{equation}
The singular fiber can be defined \cite{Kodaira} simply as the 
divisor of the holomorphic function $\tau(\phi)$, with $\tau$ a 
uniformization parameter on $\Delta$. What we will generically 
call a ``geometric'' soft breaking term would be defined exactly 
by this holomorphic function. Now we will consider the 
elliptic fibration defining the solution of the four 
dimensional $N=2$ theory from the point of view of the 
elliptic fibration defining its $N=4$ three dimensional 
compactification. The $N=2$ soft breaking term, i. e., what 
is going to play the role of $N=1$ superpotential, will 
appear as the "rotation" in an appropiated sense of the 
"geometrical" soft breaking term defined by the $N=4$ 
elliptic fibration. In order to be a bit more explicit 
we need to consider in more detail the role played by 
the different $U(1)$ entering into the game.
Let us recall that the basic dynamical fact underlying 
the instanton generation of superpotentials in three 
dimensional supersymmetric gauge theories \cite{AHW,SW3d} 
is, on one side, the generation of effective fermionic 
vertices by three dimensional instantons and, on the other 
side, the existence of a non anomalous $U(1)$. The way these 
two facts combine is thanks to the role of the dual photon 
field as a Goldstone boson and to the form of the instanton 
contribution \cite{P} as $\exp -I+i\sigma$, with $\sigma$ 
the dual photon field 
and $I$ the standard instanton action in the Prasad-Sommerfeld 
limit. For $N=4$ three dimensional theories 
the moduli space is an hyperk\"ahler manifold, that for 
the case of $SU(2)$ coincides with the Atiyah-Hitchin 
space for the moduli of two static BPS monopoles \cite{AH}. After 
distinguishing one particular complex structure, the 
hyperk\"ahler moduli space becomes an elliptic fibration. 
The rotation group $SU(2)_R$
acting on the Atiyah-Hitchin space gets effectively 
reduced to a $U(1)$ once we distinguish a particular 
complex structure. This $U(1)$ is in fact not broken 
by instanton effects in three dimensions and therefore 
we should be able to see its action directly on the 
elliptic fiber. This was explicetely done in reference 
\cite{SW3d} for the case of $SU(2)$. In fact from the Atiyah-Hitchin 
solution
\begin{equation}
y^2 = x^2v +x
\label{4}
\end{equation}
the $U(1)$ $C^*$ action on the curve is defined by:
\begin{eqnarray}
y & \rightarrow & \lambda y, \nonumber \\
x & \rightarrow & \lambda^2 x, \nonumber \\
v & \rightarrow & \lambda^{-2} v, 
\label{5}
\end{eqnarray}
As we observe from equation (\ref{5}), the $U(1)$ is acting on the 
$v$-plane. What we have called the "geometric" soft breaking 
term can be read off from the curve defined in equation
(\ref{4}); namely, we get
\begin{equation}
v = - \frac {1}{x}.
\label{6}
\end{equation}
Now we want to get from (\ref{6}) the soft breaking term of the four 
dimensional $N=2$ gauge theory. The solution for $SU(2)$ is 
defined by the following elliptic fibration:
\begin{equation}
y^2 = x^3-x^2u+x.
\label{7}
\end{equation}
The first thing to be noticed is that the $U(1)$ is on the 
$u$-plane, effectively broken to a ${\bf Z}_2$ subgroup. Moreover, 
as introduced in reference \cite{SW3d}, in order to go from (\ref{4}) 
to (\ref{7}), 
we can use the following change of variables:
\begin{equation}
v=x-u
\label{8}
\end{equation}
from which it is manifest the existence of only a ${\bf Z}_2$ subgroup 
on the $u$ plane once we asign charges $2$ and $-2$ to $v$ and 
$x$ as dictated by (\ref{5}). The soft breaking term for the $N=2$ 
theory, which is our candidate for the superpotential of the 
four dimensional $N=1$ theory will be defined by combining 
the "rotation" (\ref{8}) in the $(x,v)$ plane with the "geometric" soft 
breaking term given by equation (\ref{6}). Of course we need to include 
some scale that is effectively characterizing the breaking of 
the $U(1)$ symmetry. This scale is $\Lambda_{N=1}$, and should 
vanish in the compactified $U(1)$ invariant case:
\begin{equation}
U = \Lambda_{N=1}^6 x + \frac {1}{x}.
\label{9}
\end{equation}
This was the derivation of the superpotential in 
reference \cite{SW3d}. The crucial step is of course to 
pass from the complex structure with explicit $U(1)$ 
action, the hyperk\"ahler Atiyah-Hitchin space defined in 
(\ref{4}), to the elliptic fibration parametrized by $u$ where 
the $U(1)$ action is manifestly broken. The change of variables 
(\ref{8}) is only allowed when we take out a point in the elliptic 
fiber. This is related to the fact that the volume of the 
elliptic fibration becomes infinity in the three dimensional 
limit. Notice also that the minima of (\ref{6}), interpreted as a 
function of $x$, are at the point of infinity. This raise 
inmediately the question on what can be the meaning of the 
change of variables (\ref{8}) in the four dimensional limit. In the 
solution of the $N=2$ theory the elliptic curve degenerates 
in two points (for the case of $SU(2)$) where the cycle of 
the torus vanishes, for that curve it is still possible to 
define the change of variables (\ref{8}). In fact what we observe 
comparing (\ref{6}) with (\ref{9}) is that the minima 
at infinity of (\ref{6}) 
become now two finite points, with two well defined 
"critical values" precisely at the points 
in the $u$ plane where the curve degenerates.  

\section{$U(1)$ and M-theory instantons.}

In reference \cite{Winst} Witten has introduced M-theory 
instantons as six-cycles of aritmetic genus equal one 
in M-theory compactifications on Calabi-Yau fourfolds 
with $SU(4)$ holonomy. Let us consider a Calabi-Yau 
fourfold elliptically fibered. More specifically we 
will consider that on the base space $B$ we have a four 
dimensional locus $C$ on which the fiber degenerates in 
some of the ADE Kodaira different types. Moreover and 
following reference \cite{KV} we assume that $C$ satisfies 
the condition $h_{1,0}=h_{2,0}=0$. In order to mimic the effect 
of three dimensional instantons the first thing to do is to 
properly identify the non anomalous $U(1)$ symmetry for 
which the instanton defined by the six cycle produces the 
correct change of $U(1)$ charge. This was done in reference 
\cite{Winst}. In fact the normal bundle to the six-cycle, in our 
case the normal to the locus $C$ in the base space $B$, is 
one complex dimensional, and the desired $U(1)$ can be identify 
with
\begin{equation}
z \rightarrow e^{i\theta} z.
\label{10}
\end{equation}
By analogy with what we did in the $SU(2)$ case in the 
previous section we should now identify the coordinate 
$z$ on the normal bundle as the analog of the $v$ variable, 
as on both of them, there is a well defined $U(1)$ action
\footnote{We are effectively thinking of the 
elliptically fibered $z$-plane as the analog of the $
N=4$ moduli in three dimensions. The topological 
conditions we have impossed on $C$ seem to allow 
that picture.}. Now we need the analog of equation (\ref{6}); 
what we have called the "geometric" soft breaking term. 
The coordinate $z$ has the appropiate $U(1)$ charge to 
define a superpotential in the $N=2$ three dimensional 
theory. Let us supposse that we are dealing with a 
singularity of type $A_{n-1}$, then following \cite{KV} 
we can associate with each irreducible component a 
six-cycle defined by just fibering the irreducible 
component on the locus $C$. These are six-cycles of 
arithmetic genus equal one. We will only consider  
the irreducible components that contribute to the 
Picard group, so for a singularity of type $A_{n-1}$ we 
have $n-1$ contributions. Each one can be represented 
by a variable $1/x$ with the appropiated $U(1)$ charge, 
hence the analog of equation (\ref{6}) of the previous section 
should be
\begin{equation}
z= - \frac {n-1}{x}
\label{11}
\end{equation}
Now the rule we have abstracted from the previous 
analysis of $N=4$ three dimensional $SU(2)$ super 
Yang Mills would be -after interpreting the variable 
z in parallel to the $v$ of the previous section - to define 
the soft breaking term of the $N=2$ four dimensional theory 
i.e the superpotential of the four dimensional $N=1$ theory. 
This can be done by the equivalent to the change of variables (\ref{8}), 
which now should mix the $x$ and $z$ variables. In other 
words, we are rotating the elliptic fiber of the elliptically 
fibered Calabi-Yau fourfold to a new elliptic fibration 
variable, $u$, and the way to do it would be simply by, as 
was the case in the field theory analysis of the previous 
section, breaking the $U(1)$ symmetry this time to $Z_n$:
\begin{equation}
z=x^{n-1} -u,
\label{12}
\end{equation}
which implies for $u$, using the $U(1)$ charges (\ref{5}) for $v$ 
and $x$ a residual $Z_n$ symmetry. In fact, it follows from 
(\ref{12}) that the residual symmetry is characterized by $u
\rightarrow \lambda^2 u$ with $\lambda^{2n}=1$. 
 As before the superpotential will be now obtained 
combining (\ref{11}) and (\ref{12}) and including the scale of the $U(1)$ 
breaking,
\begin{equation}
U = \Lambda_{N=1}^{3n} x^{n-1} + \frac {n-1}{x}
\label{13}
\end{equation}
The contribution (\ref{11}) survives the decompactification 
limit since it is coming
from vertical instantons \cite{Winst}.
This superpotential has the nice features we expect 
in $N=1$ four dimensions. When we consider (\ref{12}) as a 
superpotential and we look for its ``critical values'' 
in the $u$ plane it seems like if, by going from the 
$U(1)$ invariant picture in the $z$ plane to the one 
with broken $U(1)$ in the $u$ plane, we have effectively 
"broken" the initial $A_{n-1}$ singularity into $n$ $A_0$ 
singularities, separated by the scale of the $U(1)$ 
breaking i.e $\Lambda_{N=1}$.
We would like to interpret this effective "breaking" 
of the elliptic singularity, when we move to the $u$ 
plane as reflecting a type of confinement dynamics 
based on "liquids" of fractional topological objects 
separated a distance of the order of the scale $\Lambda_{N=1}$, \cite{GH}. 
Here it is important to stress that fractional topological 
objects have nothing to do a priori with any fractionalization 
of the standard $\theta \rightarrow \theta + 2 \pi$ symmetry transformation. In fact 
if one uses torons \cite{tHcmp} as fractional topological objects to get 
the gaugino condensate, what one gets is $<\lambda \lambda>_l \simeq 
\Lambda_{N=1}^3 e^{2 \pi i l/n}$ with $l=0, \cdots,n-1 $ 
representing an electric flux flow. By changing 
$\theta \rightarrow \theta + 2 \pi$ we pass from one vacua to another as a 
direct consquence of Witten's dyon effect \cite{Wdy}. In fact the 
electric flow changes as $l + \frac {\theta \vec{m}}{2\pi}$. In reference \cite{Wlast} 
a beautifull proposal for QCD string has been made on 
the basis of a D-brane interpretation of domain walls. 
If we consider (\ref{13}) in the same spirit as a Landau-Ginzburg 
superpotential the domain wall would be characterized by the 
difference between two consecutive critical values. Now, if we 
use the toron representation of the gaugino condensate in 
terms of `t Hooft's
electric flux, the operator to transit from one vacua to 
another would be the Wilson loop, which would in principle allow the 
comparison of both quantities.  

\section{M-theory fivebranes.}

The physical picture of instantons in \cite{Winst} was that 
of fivebranes wrapping on the six cycles of the 
Calabi-Yau fourfold. Recently a description of four dimensional 
$N=1$ theories has been done using branes configurations 
and M-theory \cite{Oz,Wlast}. In that picture a rotation of branes 
\cite{barbon} was introduced in order to break supersymmetry. We 
think that that type of rotations in the brane description 
of $N=1$ four dimensional theories is equivalent to the 
change of variables (\ref{12}) suggested by the three dimensional 
case as given by (\ref{8}). Moreover the normal bundle is playing 
in this picture the analog role of the $(8,9)$-plane of 
references \cite{Oz,Wlast}, and the conditions for rotating the 
curve are the analogous to those necessary for giving 
sense to the change of variables (\ref{8}) and (\ref{12}). The hope of 
this note is by relating the superpotentials of type (\ref{12}), 
much in the spirit of reference \cite{KV}, to M-theory 
instantons, to point out to the common dynamical 
mechanism of confinement and gaugino condensates. 
On the other hand it would maybe be interesting to 
work out in more detail the connection between brane 
manipulations, as for instance rotations, and the type 
of variable changes underlying the relation between the 
elliptic fibration for three dimensional theories and 
those of their uncompactified four dimensional limit, 
mostly taking into account that they share common 
complex structures.

\vspace{20 mm}
  
\begin{center}
{\bf Acknowledgments}
\end{center}

This work is partially supported by European Community grant 
ERBFMRXCT960012, by grant AEN96-1655, and by OFES930083.

\end{document}